\documentclass{amsart}

\title{Tetrions: a discrete approach to the standard model}
\author{Robert Arnott Wilson}
\date{First draft: 20th January 2023. This version: 25th January 2023.}

\address{Queen Mary University of London}
\email{r.a.wilson@qmul.ac.uk}

\newcommand{\CC}{\mathbf C}
\newcommand{\RR}{\mathbf R}
\newcommand{\HH}{\mathbf H}
\newcommand{\rep}{\mathbf}

\begin{document}
\begin{abstract}
I show how the symmetry-breaking of a recently proposed embedding of the
standard model of particle physics in $E_8$ can be explained in terms of the representation
theory of the binary tetrahedral group. This finite group 
provides a link
between various types of spin and isospin that can be exploited to `explain' the chirality of the weak interaction,
and the existence of three generations of fermions. Two apparently small technical differences between the finite group model and the standard model
turn out to have profound consequences for the ways in which the weak and strong forces create mass.
\end{abstract}
\maketitle
\section{The mathematical model}
\subsection{Introduction}
In a recent paper \cite{MDW}, Manogue et al. propose a new $E_8$ model of particle physics, that differs from
other similar models \cite{Lisi,Lisi2,Chester} in one important respect, namely that the gauge group of the strong force
appears in its split rather than compact real form. In order to reproduce the standard model, a complex structure is
then imposed on the \emph{representations}, rather than the group itself, and it is argued that this is sufficient
for all practical purposes.
The actual group that is used is then
\begin{align}\label{thegroup}
(SL(2,\CC)\times U(1))/Z_2 \times U(1) \times SU(2) \times SL(3,\RR),
\end{align}
compared to the standard model:
\begin{align}\label{SMgroup}
SL(2,\CC) \times (U(1)\times SU(2) \times SU(3))/Z_6.
\end{align}
The technical differences in the scalars in the two cases are of no deep significance for the underlying symmetries, as they merely select
different representations of the group for use in the physical theory. But the difference between $SU(3)$ and $U(1) \times SL(3,\RR)$ is,
I shall argue in this paper, not merely technical, as proposed in \cite{MDW}, but deeply significant, and has far-reaching consequences for the understanding of mass and the quantisation of gravity.

As is usual in grand unified theories,
the key question of why the symmetries of $E_8$ break down to this group is not addressed in \cite{MDW}. 
Regardless of the actual physical mechanism of this symmetry-breaking, it is unsatisfactory from the theoretical point of
view that there is no \emph{mathematical} reason for it. A Lie algebra \cite{Zee,WoitQFT} does not naturally break up in this way. A
Clifford algebra \cite{Porteous}, often used in models of  
physics, can break up as a sum of two matrix algebras, but not more, and when it does break in this way, the two
factors are isomorphic.  Hence neither of these types of algebra can provide a convincing mathematical
explanation for the 
breaking of  
symmetry down to the standard model.

The most obvious examples of algebras that do break up naturally as direct sums of matrix algebras
are the real and complex group algebras of finite groups \cite{JamesLiebeck}.
In order to obtain the Lie group (\ref{thegroup}), in which real and complex matrix algebras both appear,
 a real group algebra is required. Moreover, the finite group must have irreducible representations of complex
dimensions $1$ and $2$, and another of real dimension $3$. 
Assuming that at least one of the $2$-dimensional representations is faithful, there are
essentially just 
three possible groups \cite{Blichfeldt,Klein1,DuVal}, namely the binary tetrahedral, binary octahedral and binary icosahedral groups, of orders
$24$, $48$ and $120$ respectively.  These are precisely the three finite subgroups of $SU(2)$ than act irreducibly on the
spin $1$ (three-dimensional) representation, but they can also be modified by arbitrary mixing with finite groups of complex scalars.

The representation theory of these groups
is studied in detail in \cite{finite,octahedral,gl23model,icosa,ZYM} with a view to
using the group algebras in potential new models of fundamental physics. The arbitrary scalars have minimal effect on the representation theory,
so they can be omitted without any real loss to the model, and may not contribute anything tangible to physics.
In this paper I present  an updated version of the tetrahedral case \cite{finite}, and explain how it can be related to the
$E_8$ model \cite{MDW} and to the standard model.

\subsection{The binary tetrahedral group}
It is worth remarking at this point that the use of the binary tetrahedral group as a finite version of the weak gauge group $SU(2)$ goes back
to the original work of Yang and others \cite{Yang} from the 1950s. More recent work \cite{Frampton1,Frampton2,Frampton3} shows that this
use of the finite group has real predictive power for the measured values of mixing angles in the standard model.
It is therefore of interest to try to turn this phenomenological approach into a deeper theoretical basis for the properties of
neutrinos and quarks, in order to explore where the various masses and mixing angles ultimately come from.
Where the present paper goes beyond these earlier papers is in proposing a theoretical basis in the binary tetrahedral group,
not only for the weak force, but also for the strong force and electromagnetism.

It turns out, and is well-known \cite{finite}, that 
the real group algebra of the 
binary tetrahedral group is isomorphic to the
following sum of full matrix algebras
\begin{align}
\RR^{1\times1}+\CC^{1\times1}+\RR^{3\times3}+\HH^{1\times1}+\CC^{2\times2},
\end{align}
where $\HH$ denotes the algebra of Hamiltonian quaternions.
By taking the matrices of \emph{real} determinant $1$ in each factor, we obtain exactly the group
(\ref{thegroup}). This remarkable coincidence demands some explanation. But it is possible that it is
purely a coincidence, driven by the intuition of the authors of \cite{MDW},
rather than any real mathematical or physical connection between the group algebra and the Lie algebra.
If so, then the binary tetrahedral group might be capable of providing a much more fundamental
explanation for the standard model than any $E_8$ model could ever do. 

It is, of course, possible to derive the standard model of particle physics from the group algebra by following the recipe
in \cite{MDW}. However, it should be clear that the structure of $E_8$ itself plays only a small part in this construction.
It will therefore be enlightening, if possible, to remove this complicating step, and derive the standard model directly from the
representation theory of the binary tetrahedral group. 

The finite group itself provides an extra ingredient that is not in
any Lie group model, and enables us to link the different gauge groups together in order to `explain' the observed `mixing' of the
different fundamental forces at the quantum level, but without introducing any new unobserved degrees of freedom.
In particular, it turns out that the somewhat \emph{ad hoc} formalism of electro-weak mixing in the standard model can be explained
by the fact that two different spinor representations of the finite group are treated as if they were the same. Separating these two
representations again enables us to revise the formalism in such a way as to incorporate the three generations of fermions.

\subsection{Real and complex representations}
The binary tetrahedral group is most easily described as the multiplicative group of the $24$ quaternions
\begin{align}
\pm1, \pm i, \pm j, \pm k, (\pm1\pm i\pm j\pm k)/2.
\end{align}
It consists of seven conjugacy classes of elements, represented by $1$, $-1$, $i$ and $(\pm1 \pm i+j+k)/2$. It therefore \cite{JamesLiebeck}
has seven
complex irreducible representations, whose characters (giving the traces of the representing matrices) are the rows in the following table,
where $w$ denotes the quaternion $(-1+i+j+k)/2$ and $\omega$ denotes the complex number 
$(-1+\sqrt{-3})/2$.
\begin{align}
\begin{array}{l|cccc}
& \pm1 & i & \pm w & \pm w^2\cr\hline
\rep1a & 1 & 1 & 1 & 1\cr
\rep1b & 1 & 1 & \omega & \bar\omega\cr
\rep1c & 1  & 1 & \bar\omega & \omega\cr
\rep3 & 3 & -1 & 0 & 0\cr
\rep2a & \pm2 & 0 & \mp1 & \mp1\cr
\rep2b & \pm2 & 0 & \mp\omega & \mp\bar\omega\cr
\rep2c & \pm2 & 0 & \mp\bar\omega & \mp\omega\cr\hline
\end{array}
\end{align}
Of these, $\rep1a$ and $\rep3$ are real, $\rep2a$ is pseudoreal, that is a one-dimensional quaternionic representation, and the rest
are genuinely complex.

The characters of the real representations are therefore as follows:
\begin{align}
\begin{array}{l|ccc}
&\pm1 & i & \pm w\cr\hline
\rep1 & 1 & 1 & 1\cr
\rep2 & 2 & 2 & -1\cr
\rep3 & 3 & -1 & 0\cr
\rep4a & \pm4 & 0 & \mp2\cr
\rep4b & \pm4 & 0 &  \pm1\cr\hline
\end{array}
\end{align}
While the standard model always works with complex representations, real physical measurements are always expressed as real numbers,
so that the real representations may ultimately provide a better model. For present purposes, we work mainly with the
complex representations, since they are more familiar and also translate more easily into the standard model.

\section{Deriving the standard model}
\subsection{Spinors and isospinors}
The complex representations $\rep2b$ and $\rep2c$ are complex conjugates of each other, acted on by $SL(2,\CC)$, and
therefore correspond to left-handed and right-handed Weyl spinors in the standard model. The complex representation $\rep2a$,
on the other hand, is acted on by $SU(2)$, as the gauge group of the weak force, and therefore corresponds to (weak) isospin.
There is no natural action of this gauge group on the left-handed Weyl spinor, but the finite group acts on both spinors and isospinors
simultaneously, so can be used to transfer some approximate action of $SU(2)$, or 
a subgroup thereof, from one to the other. 

For example, the tensor product
\begin{align}
\rep1b \otimes \rep2a = \rep2b
\end{align}
could be used to identify $\rep2a$ with $\rep2b$ as `the' left-handed spinor. In particular, this has the effect that the 
copy of $U(1)$ that acts on $\rep1b/\rep1c$ and $\rep2$ is identified with the copy of $U(1)$ that acts
as complex scalars on $\rep2b/\rep2c$ and $\rep4b$, thereby reducing to a single copy of $U(1)$, as in the
standard model. 
Alternatively, one can turn this around, and use only the
representations $\rep1b$ and $\rep2b$ explicitly, leaving $\rep2a$ implicit. 
This is in effect what the standard model does, using the complex representation $\rep1b$ to implement $1-\gamma_5$ acting on the
left-handed spinor $\rep2b$. In this way, only the first component of isospin appears explicitly in the model, and the `direction' of this component,
like the `direction' of spin, is chosen arbitrarily, without regard to the geometry of the outside world.

This procedure has the effect of incorporating the $120^\circ$ rotation $\omega$ into the structure of the spinor, in place of the discrete
symmetry $w$. Compared to the standard multiplication by $i$, that is a $90^\circ$ rotation, to get $\gamma_5$ from $-i\gamma_5$,
there is a discrepancy of $30^\circ$.
It is plausible that this angle is
a first approximation to the electro-weak mixing angle (Weinberg angle), that does not take into account the running
with the energy scale. In order to include the running, it is necessary to include the real scalar $\rep1a$ as well as the complex pseudoscalar 
$\rep1b$, so that  a single angle is no longer sufficient to describe the full structure of the mixing. 

At the same time, the implicit identification of the real representation $\rep1a$ with the imaginary part of 
the complex representation $\rep1b$ destroys the discrete symmetry of order $3$, which may explain why the standard model
loses the generation symmetry at this point, and has to insert it again by hand later on.
In this paper, I propose to reinstate the generation symmetry in such a way that it corresponds to permutations of the three
`directions' of isospin.
This contrasts with the standard model, in which the direction of isospin is not explicitly used, and all three generations use the same 
arbitrary direction.

When we convert to the real representation, 
the Dirac spinor $\rep2b+\rep2c$ reduces to a Majorana spinor $\rep4b$,
while the isospinor $\rep2a$ becomes $\rep4a$. It is no longer possible to separate $\rep2b$ from $\rep2c$ as left-handed and
right-handed spinors, but the separation of isospinors $\rep4a$ from spinors $\rep4b$ is clear enough. On the other hand, if
we lose the generation symmetry as above, then there is no distinction between the representations $\rep4a$ and $\rep4b$,
so that the distinction between spin and isospin becomes less clear. It would seem therefore that
the best way to relate the finite group algebra to the standard model would be
to adopt a mixed strategy, working with the real group algebra, but the complex representations.

The conclusion we must surely draw from this discussion is that although the standard model can be derived from this group algebra in the way
it is done in \cite{MDW}, we lose the generation symmetry by doing so, and we lose the distinctions between the different
representations. Indeed, we lose all the symmetry, since $w$, which does not commute with $i$,  has been replaced by the scalar $\omega$,
which does.
Hence
the finite group itself plays no role in the standard model. This suggests that by keeping the finite group
we may gain some insight into \emph{why} there are three generations of fermions, and \emph{why} the weak interaction is `chiral'.
Moreover, since the finite group links the `direction' of spin to the `direction' of isospin, we may gain some insight into
\emph{how} the physical mechanisms of entanglement work.

\subsection{Tensor products}
The Dirac algebra (complex Clifford algebra) used in the standard model is, as a representation of the Lorentz group, simply the
square of the Dirac spinor. In order to identify it as a representation of the finite group, we must calculate:
\begin{align}
\rep2a\otimes \rep2a = \rep2b\otimes \rep2c & = \rep1a+\rep3,\cr
\rep2b\otimes \rep2b = \rep2a\otimes\rep2c & = \rep1c+\rep3,\cr
\rep2c\otimes\rep2c = \rep2a\otimes\rep2b & = \rep1b+\rep3.
\end{align}
If we regard $\rep2b+\rep2c$ as the Dirac spinor, as is required by the interpretation of $SL(2,\CC)$ as the Lorentz group,
then the spacetime representation is the real $4$-dimensional
representation $\rep1a+\rep3$. But if we use the above identification of $\rep2a$ with $\rep2b$, then spacetime becomes $\rep1c+\rep3$,
with a complex time coordinate, rather like an anti-de Sitter spacetime.

The Dirac equation can be obtained by interpreting $\rep1a$ as time/energy, $\rep3$ as space/momentum, and the imaginary
part of $\rep1b/\rep1c$ as mass. In discrete form, this equation comes from the tensor products
\begin{align}
(\rep1a+\rep3)\otimes (\rep2b+\rep2c) & = \rep2a^2+\rep2b^3+\rep2c^3,\cr
(\rep1b+\rep1c)\otimes (\rep2b+\rep2c)  & = \rep2a^2+\rep2b+\rep2c,
\end{align}
via a complicated process of projecting the right-handed sides onto a Dirac spinor.
For a massive fermion at rest, this involves identifying the two spinors
\begin{align}
\rep1a\otimes (\rep2b+\rep2c) & = \rep2b+\rep2c,\cr
\rep1b \otimes (\rep2b+\rep2c) & = \rep2c+\rep2a.
\end{align}
Such a procedure can of course only be carried out if we ignore the finite symmetry group, and ignore the weak interaction.

 In particular, it should be noted that the action of $SO(3,1)$ on spacetime (or its dual) cannot be realised in the
group algebra, since the latter does not contain a subgroup $SO(3,1)$. It is not possible to implement Lorentz transformations that mix space with time, but it \emph{is} possible to implement
time dilation by scalars on $\rep1a$, and it is possible to implement space contraction in arbitrary directions by elements of 
$GL(3,\RR)$. It is therefore possible to model the practical effects of Lorentz transformations to a large extent. 

If, on the other hand, we keep the finite symmetry group, 
then we need to consider all three of $\rep1a$, $\rep1b$ and $\rep1c$ as different versions of `time',
appropriate for different parts of the theory. The sum $\rep1a+\rep1b+\rep1c$ is the regular representation of the cyclic group of order $3$,
and therefore is appropriate for a discrete triplet symmetry.
This might, for example, be the generation symmetry, or perhaps it is more likely to be the broken symmetry
of the intermediate vector bosons, $Z^0$, $W^+$ and $W^-$. The former suggests a possible interpretation of the three 
generations as a three-dimensional `time', as is done explicitly by Chester et al. 
in \cite{Chester}, again in the context of $E_8$. However, the group algebra
model gives no support to the idea that there is a symmetry group $SO(3,3)$ of 
spacetime, even locally. What appears to happen is that the Euclidean
metric on space disappears, and the group $SO(3,3)$ reduces to $SL(3,\RR)$, which 
seems to indicate that
the metric structure of spacetime breaks down at a sufficiently small scale. In Section~\ref{triplets} of
this paper I use this group to propose instead a quite different implementation
of the three generations.

The corresponding calculations with the real representations are
\begin{align}
\rep4a\otimes\rep4a & = (\rep1+\rep1+\rep1+\rep3) + (\rep3+\rep3+\rep3+\rep1)\cr
\rep4a\otimes\rep4b & = \rep2+\rep2+\rep3+\rep3+\rep3+\rep3\cr
\rep4b\otimes\rep4b & = (\rep1+\rep2+\rep3) + (\rep3+\rep3+\rep3+\rep1)
\end{align}
However, the natural interpretation of $\rep4a$ is as a one-dimensional quaternionic representation, so that we should really calculate
the quaternionic tensor product, which is a real four-dimensional representation:
\begin{align}
\rep4a\otimes_\HH \rep4a = \rep1+\rep3.
\end{align}
Since $\rep4b$ is naturally complex, the tensor product $\rep4a\otimes \rep4b$ can only be real, as above, or complex $\rep2a\otimes \rep2b$, which requires breaking the symmetry of the
quaternions, that is, breaking the symmetry of the weak $SU(2)$. In other words, a \emph{complex} model of electro-weak unification
is forced to break the $SU(2)$ symmetry, but a \emph{real} or \emph{complex quaternionic} model is not. 
In particular, a model of the latter type could in principle implement this symmetry-breaking as contingent, rather
than absolute, and might therefore suggest a (physical, rather than mathematical) reason \emph{why} this symmetry-breaking
happens the way it does in the part of the universe that we can directly measure.

\subsection{Triplet symmetries}
\label{triplets}
The only representation left to consider is $\rep3$, which supports a group $SL(3,\RR)$ that we suppose must be related to the
gauge group $SU(3)$ of the strong force. Of its eight degrees of freedom, only three are compact, and can therefore be plausibly quantised.
The other five cannot be quantised in any obvious way, 
but must nevertheless relate to properties of quantised (elementary) particles. The only plausible
parameters of this type are masses, but there may be some freedom to 
choose which five masses are regarded as fundamental. The representation theory gives us
\begin{align}
\rep3\otimes\rep3 = \rep1a+\rep1b+\rep1c+\rep3+\rep3
\end{align}
which implies that there is a triplet of masses and two singlets. A possible choice may be three generations of electrons, together with a proton
and a neutron, although there are certainly other possibilities.

We should also note that the finite group links the $SO(3)$ symmetry of space to an $SO(3)$ subgroup of the $SU(3)$ 
colour symmetry of quantum chromodynamics (QCD).
It is not clear exactly how to interpret this physically, although it seems clear that the `colours' must be related to some kind of `orientation' in space.
But this is all happening on such a small scale that it may not be possible to resolve such questions experimentally.

Similarly, the triplet (flavour) symmetries of neutrinos can only plausibly be derived from the square of the isospinor $\rep2a$, that is
$\rep1a+\rep3$, 
so must be implemented in $\rep3$, and not in $\rep1a+\rep1b+\rep1c$.
On a macroscopic scale relevant for the study of neutrino oscillations, the representation $\rep3$
has interpretations as space, momentum, and various classical fields including electric, magnetic and gravitational.
It is possible,
therefore, that the generation of a neutrino that is detected by a microscopic experiment using the finite group of symmetries
of $\rep3$ can change when the ambient spacetime coordinates change, due to rotations or other accelerations, or to a gravitational field.

The simplest explanation of
neutrino oscillations in this context is  
that $\rep3$ contains the momentum, measured with respect to the symmetry group $SO(3)$ of the ambient space,
and that $\rep1a$ contains the energy, measured with respect to the speed of light. There is then no room for an intrinsic mass, but there is
room for an effective change in mass due to bending of the path of the neutrino in a variable gravitational field, or due to the acceleration of
one observer with respect to another. In contrast, the standard model effectively identifies $\rep1a$ (here identified as
the energy) with the imaginary part of $\rep1b$
(the Dirac mass), which forces the neutrinos to have a non-zero (Dirac) mass. 
But general relativity does not require a Dirac mass in order for the neutrinos to be affected by the force of gravity,
possibly even leading to a change of flavour in an appropriate quantum-gravitational context.

\subsection{The strong force}
As complex tensor products of complex representations, we have 
\begin{align}
\rep1b\otimes \rep3 = \rep1c\otimes \rep3 & = \rep3,
\end{align}
but there is a subtlety here caused by the fact that the factor $\rep3$ in the tensor products can be real,
whereas the resulting $\rep3$ on the right-hand side cannot. Nevertheless, a (real) basis can be chosen such that
the latter splits as the sum of two real representations. 
If we extend to Lie groups acting on the two factors
\emph{independently}, then this splitting can still be achieved, 
giving rise to a group $U(1) \times SL(3,\RR)$ with compact part $U(1)\times SO(3)$, in which $U(1)$ acts as scalars.
But this is not what happens, either 
in the standard model, or in physical reality.
There is a `mixing' between the $U(1)$ factor and a copy of $SO(2)$ inside $SO(3)$ caused by the fact that an element of order $3$
in the finite group acts on both simultaneously.

What then happens in the standard model is that there is no clean relationship between the bases with respect to which the two
groups $U(1)$ and $SO(3)$ are written. Moreover, since $U(1)$ does not act as a group of scalars,
these two groups no longer generate a direct product $U(1)\times SO(3)$, and instead generate $SU(3)$. The theory of QCD is then built on
this basis, and although it works well, it appears to be built on rather shaky foundations. If the group algebra is the fundamental structure
underlying the standard model, then 
the fact  that  this algebra does not contain a copy of $SU(3)$ is important. 
In this case, we must conclude that (at least) five of the eight `gluons'
of QCD are entirely fictitious concepts: not only are they unobservable, but they have no physical reality at all. 
Controversial as this may sound, it is not 
a new idea: it is already implicit (though not explicit) in \cite{MDW}, since there can be at most three gauge bosons arising from the
compact part of $SL(3,\RR)$. Since direct model-independent tests of properties of gluons are virtually impossible,
this idea cannot be easily falsified or rejected.

Nevertheless, the parameters that are required to change basis between the three isomorphic complex representations
$\rep1a\otimes_\CC\rep3$, $\rep1b\otimes_\CC\rep3$ and $\rep1c\otimes_\CC\rep3$ fill two complex $3\times 3$ matrices,
represented in the standard model by the Cabibbo--Kobayashi--Maskawa (CKM) \cite{Cabibbo,KM}
and Pontecorvo--Maki--Nakagata--Sakawa (PMNS) \cite{Pontecorvo,MNS} matrices.
Each contains only four independent real parameters, because the only physical part is a map between two copies of $U(1)$, and a map between
two copies of $SO(3)$. The rest of $SU(3)$ does not contain physically meaningful parameters.

\section{Beyond the standard model}
\subsection{Hidden variables}
The group algebra described in this paper is just about big enough to describe the standard model, with nothing left over.
By using the finite symmetries, we obtain three generations of fermions, which are characterised by three orthogonal components of 
a certain kind of isospin,
where only one is used in the standard model. I have not chosen a basis for this isospin, which is not necessarily identical
to the standard model weak isospin, and therefore have not chosen explicit quantum
numbers for the different particles, but in principle this should not be too difficult. Similarly in the case of spin, we are able to use all
three components in the model, rather than having to choose one. This allows an elementary particle to carry three spin quantum numbers
rather than just one, although it is probable that only two of the three are independent. 
However, it is still the case that a measurement requires a macroscopic choice
of this direction, in order to define the eigenstates, so that one can only measure spin in a specific direction, and not in two directions simultaneously.

The effect of this combination of properties is that an elementary particle carries more information than it is possible to measure.
This seemingly contradictory  
state of affairs is exactly what one needs to explain the puzzling aspects of measurements of properties of
widely-separated entangled particles---both particles carry hidden variables that are \emph{discrete}, and cannot be measured directly.
The discreteness is crucial here, as it is what enables the model to evade the well-known problems with continuous hidden variable theories
\cite{Bell,Rae,Kochen,FWT,FWT2}.
What makes the experimental properties of quantum mechanics so puzzling, and counterintuitive, is that
the hidden variables can only be detected by elementary particles, and not by macroscopic apparatus.
Since the elementary particles have only a finite number of quantum states, they can only carry a finite amount of information, and
can only distinguish a finite number of directions in space. 

This 
process 
naturally gives rise to the effect that is known in  quantum mechanics as 
`collapse of the
wave-function'. In the group algebra model, the wave-function itself is not associated to the individual particle, but to the macroscopic environment,
consisting of enough elementary particles to make it effectively continuous. The `collapse' happens when the particle interacts with the environment,
via an interaction with an individual particle in the environment. The Born probabilities arise from the fact that the experimenter has
little or no control over the
quantum properties of the environment.

\subsection{Mixing}
The finite group implements some strange mixing between the direction of spin and the direction of isospin. While the two directions certainly
appear to be independent,
they cannot apparently \emph{change} independently, since every element of the finite group acts on both
simultaneously. This fundamental property 
seems to underlie 
the observed chirality of the weak interaction. 
In the classic Wu experiment, 
a fixed direction of isospin is determined by the choice to observe the physical process of
beta decay of a neutron, which breaks the generation symmetry, and a fixed direction of spin was chosen
by the experimenters \cite{Wu}. 
A third direction was observed, namely the direction of spin/momentum of the ejected antineutrino, and was found to be not independent of the pair of choices made. Thus what might appear at first sight to be three independent copies of $\rep3$ actually contains only two.
Once two independent choices of direction have been made, the third is determined.

A similar but even more mysterious mixing seems to underlie the phenomenon of neutrino oscillation.
Here the experiments choose a particular generation of neutrino at both source and sink, and a particular
direction of momentum between them. The natural expectation would be that the first two choices must be the
same, but are independent of the third. But this is not what the experiments reveal. Yes, there are two independent
choices, rather than three, but the relationship between them is 
more subtle than would naively be expected. There is a mixing between spin/momentum on the one hand and generation/flavour
on the other. However, this is mixing is not apparent if one regards the momentum as fixed, so that a third copy of $\rep3$ is required
to explain the observed effects.

The group algebra does indeed contain a third copy of $\rep3$, which we can use if only we can identify what it is and how to use it. 
Since the experiments are large-scale, this copy of $\rep3$ must have a plausible macroscopic interpretation.
Since neutrinos do not participate in electromagnetic interactions, the only available macroscopic
interaction that is incontrovertibly known to science is gravitational. Assuming we do not wish to predict
new forces at this point, we must
then use the third copy of $\rep3$ to choose the direction of the gravitational field.
If this analysis is correct, then the probabilities of detecting neutrinos of particular flavours in particular experiments
can be predicted from the geometry of the experiment relative to the gravitational field. In particular, if there is no correlation between
the directions of the gravitational field at source and sink, as for example in solar neutrino experiments, then we should expect a $1/3$ probability
of detecting any specific flavour, as is indeed confirmed by experiment \cite{SNO}.

\subsection{Implications for quantum gravity}
The essential part of any such extension of the standard model to include gravity
is a separation of the representations $\rep2a$ and $\rep2b$, which are
mixed together in the standard model. In other words we must separate the neutrino representation $\rep2a$ (without time, and therefore without intrinsic mass) from the matter 
(Dirac spinor) representation $\rep2b+\rep2c$ (with time and intrinsic mass). The Newtonian gravitational field then appears as the symmetric square of the neutrino field. In particular, the known fact that neutrinos interact very weakly with matter,
combined with the squaring process, reduces the strength of the Newtonian gravity of an elementary particle to an undetectable level 
compared with the other forces. 

Putting a Dirac spinor in a gravitational field, we obtain the tensor products
\begin{align}
\rep2b\otimes\rep3 & = \rep2a+\rep2b+\rep2c\cr
\rep2c\otimes\rep3 & = \rep2a+\rep2b+\rep2c,
\end{align}
which not only causes a mixing between left-handed $\rep2b$ and right-handed $\rep2c$,
but also causes the Dirac particle to emit neutrinos from $\rep2a$. But it would appear from this discussion that gravitational effects
only arise from the interactions of neutrinos with each other: since such interactions have no direct effect on matter, the obvious interpretation
at a macroscopic level is that these undetectable neutrino interactions instead affect the shape of the underlying spacetime, as in the
modern textbook interpretations of general relativity. On the other hand, a hypothetical future theory that models these neutrino interactions
explicitly
might in principle be able to
explain gravity as an indirect effect of neutrino-neutrino interactions on the average effect of neutrino-matter interactions.
If so, then `curvature of spacetime' will become a redundant concept, no longer required for an explanation of gravity.

\subsection{Comparison with general relativity}
A theory of quantum gravity based on these ideas cannot be relativistic in the usual sense, 
because it is not acted on by the Lorentz group $SL(2,\CC)$.
Its gauge group is therefore not $GL(4,\RR)$, as is generally assumed \cite{GL4R1,GL4R2}, 
but only $GL(3,\RR)$, or rather, the positive determinant component
$GL(3,\RR)^+$. However, the latter group can act on $4$-dimensional spacetime as a group of matrices of determinant $1$, in which
there are elements acting as time dilation on the (fixed) time coordinate, and simultaneously as length contraction on an arbitrary space coordinate.
These are not Lorentz transformations, but their effects on measurable quantities may be almost indistinguishable from  
the effects of Lorentz transformations.

The quantisation of this gauge group forms the representation $\rep1+\rep2+\rep3+\rep3$ of the binary tetrahedral group,
while the symmetric rank $2$ tensors of general relativity reduce to
the form
\begin{align}
S^2(\rep1+\rep3) = \rep1+\rep3+\rep1+\rep2+\rep3.
\end{align}
Removing the Ricci scalar from the Ricci tensor to leave the Einstein tensor shows that the gauge group $GL(3,\RR)$ accurately
quantises the Einstein tensor. But there is no $SO(3,1)$, so this gives us a \emph{non-relativistic} version of general relativity, if that isn't an oxymoron.
In other words this quantum gravity is indistinguishable from general relativity in the non-relativistic limit.
However, this quantisation procedure 
does not exhaust the group algebra, which contains an extra vector field $\rep3$ compared to the Einstein tensor.

Since the entire bosonic part of the group algebra can be expected to have detectable effects on all scales, including galactic and cosmological scales,
we must do three things to extend general relativity to a complete theory of gravity: first, remove the Ricci scalar; second, add a new vector field;
and third, remove the requirement for invariance under $SO(3,1)$.
Various models of gravity have been developed along these lines, of which the most successful to date are those that add a vector and one or two
scalars to the symmetric tensor of general relativity \cite{Sanders,TeVeS,AeST}. But none of these removes the $SO(3,1)$ symmetry,
and therefore, if my analysis is correct, they are all inconsistent with quantum mechanics. 

In the model proposed here,
Lorentz invariance is only an approximate symmetry of the macroscopic world, valid in regions of the universe in which there is a
reasonably well-defined concept of inertial frame. Presumably then, while an approximately consistent definition of inertial frame can be 
made on a Solar System scale, provided that curvature of spacetime is invoked to compensate for the gravitational fields of the objects within the
Solar System,
the same is not necessarily true on a galactic scale.

It should be noted also that the separation of $\rep2a$ from $\rep2b+\rep2c$ entails a separation of $\rep1a$ from $\rep1b+\rep1c$, 
and therefore a separation of
a real scalar energy from a complex scalar mass plane. This entails adding an extra scalar field to the usual mass and energy fields. There are various possible ways to interpret this extra scalar, but its practical effect is to separate the concepts of inertial and gravitational mass.
While the local experimental evidence is that these two concepts track each other closely, the non-local astronomical evidence is that they
do not \cite{Milgrom1,Milgrom2,Milgrom3,MOND,MOND2,Banik,newparadigm}. 
The model proposed in this paper therefore provides a potential way to reconcile these two conflicting strands of evidence.

\subsection{Physical significance of a non-compact gauge group}
In \cite{MDW}, the change from the compact gauge group $SU(3)$ to the split group $SL(3,\RR)$ is presented as a mere technicality,
since the calculations required in the standard model can always be patched up by introducing factors of $i$ from the group $U(1)$.
To a certain extent this is true, at least from a practical point of view. 
It is much the same process that is already incorporated into the standard model in the Dirac algebra, where the complexification allows
one to switch between the Lorentz group in the form $SL(2,\CC)=Spin(3,1)$ and the compact form $SU(2)\times SU(2)=Spin(4)$.

But in terms of the underlying conceptual structure, the difference between a Lorentzian (Minkowski) spacetime and a
Euclidean (quantum) spacetime is important. Similarly, the difference between $SU(3)$ and $SL(3,\RR)$ for describing the strong force is
important. The split form is not compatible with
the Yang--Mills paradigm of particle physics. 
The basic problem is that compactness is required for (second) quantisation to produce identifiable gauge bosons that can be
detected in experiments. Non-compact degrees of freedom cannot be quantised into particles. 

On the other hand, the gauge bosons of the strong force
(gluons) in the standard model cannot be experimentally detected as individual particles, which implies that there is no
\emph{experimental} (as opposed to theoretical) reason why it should be necessary to quantise all $8$ degrees of freedom.
It is certainly clear that at least three degrees of freedom must be quantised, in order for the three quarks in a proton to occupy
orthogonal eigenstates, as required by the Pauli exclusion principle. 
We must conclude therefore that the strong \emph{theoretical} arguments
against a split gauge group $SL(3,\RR)$ for the strong force are not in fact supported by experiment.

It is useful at this point to 
compare with the non-compact Lorentz group in the form $SO(3,1)$, which is used in special relativity
in order to define a concept of rest mass, abstracted from the measurable quantities of energy and momentum.
Assuming an exact $SO(3)$ symmetry of space, this defines a single mass value for any particular object.
But if the symmetry of space is broken by a strongly directional gravitational field (for example), then one
can define two different mass values, one in the direction of the gravitational field
(usually called the gravitational mass), and the other perpendicular to it (usually called the inertial mass).
There is nothing in any mathematical model to say that these two masses have to be the same, 
although the experimental evidence is that \emph{locally} they are indistinguishable. The assumption that they are 
\emph{globally} the same is not justified either by experiment or by theory, but is purely a phenomenological assumption.

An analogous thing happens
in the quantum form $SL(2,\CC)$, if we consider the compact part $SU(2)$ as the gauge group of the weak force,
in order to define the intermediate vector bosons ($Z^0$, $W^+$ and $W^-$), and we use the non-compact part to
define the masses of these three particles. In this case, the symmetry-breaking is provided by the electromagnetic
force rather than the gravitational force, but the effect is the same, namely to produce two masses rather than one.
The charge conjugation symmetry prevents a splitting into three distinct masses, so that we have only one dimensionless
parameter to explain, namely the ratio of the masses of the $W$ and $Z$ bosons. While the standard model assumes that 
this mass ratio is a universal constant, the experimental evidence \cite{WZ} 
currently provides a $7\sigma$ signal that this is not in fact the case.
The group algebra model suggests that this problem might be resolved by providing greater clarity as to what precise mixture of
inertial and gravitational masses is actually being measured by the experiments \cite{toy},

Now let us consider the group $SL(3,\RR)$. Here we have five non-compact degrees of freedom, reduced to three
by the action of $U(1)$. We therefore have three independent mass values available. The representation in which
these mass values lie is the spin $2$ representation of $SO(3)$, which suggests that the underlying symmetry-breaking
comes not from the gravitational field itself, but from the tidal effects of rotations within a gravitational field (or,
equivalently, of a rotating gravitational field). I will not speculate here on which particle masses are most
appropriate here, apart from noting that the measured values of such masses might change if the tidal effects
of the rotation of the Earth relative to the the gravitational fields of the Sun and/or the Moon have a noticeable effect.
Such effects are certainly small, but they may still be detectable.
It is conceivable, for example, that the above-mentioned reported mass anomaly for the $W/Z$ bosons might have its origin in such an
effect, if one takes into account the experimental `mixing' of the weak and strong forces.

\section{Conclusion}
\subsection{Summary}
In this paper I have taken the (coincidental) isomorphism between
\begin{enumerate}
\item the group used in the `octions' model \cite{MDW} in $E_8$ as a modified version of the combined Lorentz group and gauge groups 
of the standard model, and
\item the group obtained from the determinant $1$ subgroups of the Wedderburn components of the real group algebra of the 
binary tetrahedral group \cite{finite}
\end{enumerate}
as motivation to try to use the latter algebra instead of the former as a foundation for (a modified version of) the standard model,
in an attempt to explain things that the standard model does not itself explain. 

I have shown that the apparently minor technical modifications
proposed in \cite{MDW} in fact entail profound conceptual changes compared to 
the standard model, while leaving almost all of the detailed predictions of the standard model intact.
The points of difference concern only neutrinos and gluons, and therefore have almost no effect on
any of the predictions of the standard model. They do however permit the modified model to 
predict (or postdict, or explain) some things that were not predicted in advance by the standard model, but have been detected
by experiment. These necessarily involve subtle properties of neutrinos and/or gluons,
and include neutrino oscillations, the chirality of the weak force,
the existence of three generations of fermions, the mixing of the gauge groups, and possibly even the W/Z mass anomaly.

Some of these predictions depend on an identification of part of the model as representing a quantised (Newtonian, i.e. non-relativistic) gravity.
Because it is non-relativistic, it does not satisfy all of the properties that are predicted for quantum gravity by starting from
the general theory of relativity. Nevertheless, it contains a quantised version of the Einstein tensor, and therefore contains corrections
to Newtonian gravity, that are likely to agree with general relativity to first order locally, for example on the Solar System scale.
The reason for making this assertion is that these corrections take account of accelerations, for example rotations,
but not of relativistic velocities.
At the same time, the model predicts a mixing of quantum gravity with the other forces, although I have not attempted in this paper
to make any detailed predictions of measurable effects that could be attributed to such a cause. 

\subsection{Speculations and further work}
The basic principle of the mixing of gravity with the other forces is a mixing of
neutrino generations with the direction of the gravitational field, which then leaks out via the weak force into a mixing
of electron generations with gravity. This does not necessarily affect all three generations, but it does affect at least two of them.
It therefore predicts that sufficiently precise experiments with muons in a varying gravitational field will eventually detect an
anomaly caused by a mixing of electromagnetism with gravity. 

It is possible, in fact, that this has already happened:
experimental measurement of the muon gyromagnetic ratio is in conflict with some (but not all) calculations of the standard
model prediction. The magnitude of the discrepancy in the various calculated and measured values 
\cite{muontheory,muonHVP,Fodor,muong-2} of $g-2$ happens to be
more or less equal to the sine of the angle by which the direction of the gravitational field changes over the experiment.
This might be a coincidence, or it might not: with further work it may also turn out to be a prediction of the model.

Similarly, one should expect to be able to detect mixing between gravity and the weak force by anomalies that
involve strange quarks, investigated via experiments \cite{CPexp,kaonanomaly,Kaon2}
with (neutral) kaons.
The kind of effects that might be attributable to such a cause include switching between the long and short decay eigenstates,
via interactions with (low energy, undetectable) neutrinos.
Again, the magnitude of the effect, mixing two-pion and three-pion decays, 
that was detected \cite{CPexp} in 1964 is consistent with the value of the sine of the angle by which the direction of the
gravitational field changes over the size of the experiment.
More recent experiments with rare one-pion decays \cite{kaonanomaly,Kaon2} exhibit further anomalies that might be
attributable to the same cause. Extending to the third generation one should expect to see anomalies of the same kind exhibited
by $B$-mesons \cite{Banomaly}.

Evidence for mixing of the strong force with gravity is rather harder to find, but it is possible that an inconsistency between two
different mass formulas \cite{Koide,remarks} for the mass of the tau particle might be attributable to the fact that one of these formulas
uses the proton and neutron masses, and therefore involves the strong force, whereas the other one does not. The predictions in the two cases 
differ by a proportion of less than $10^{-4}$, and both are consistent with experiment, but not with each other. The formula in \cite{Koide}
relates only the three generations of electrons, so presumably depends only on neutrinos and quantum gravity, so that it represents a 
prediction of gravitational mass. The formula in \cite{remarks}, on the other hand, involves also the proton and neutron masses, so depends
on electromagnetism, so that it represents a prediction of inertial mass. Since the model does not include an assumption that gravitational
and inertial mass are equal, it is able, at least in principle, to accommodate both of these predictions simultaneously.
Further work will be required to see whether it can in fact make either or both of these predictions.

\subsection{Evaluation}
The model presented in this paper is not a Theory of Everything (TOE). It is not a Grand Unified Theory (GUT).
It adds no new fields, no new particles, no new forces, no spin $2$ gravitons, no supersymmetry, no strings,
no new physics at all, to what is already known. All it does is explain
how the Standard Model arises from something more fundamental. In doing so, it embeds the Standard Model in the
gravitational background. This background is relativistic only because the Standard Model is relativistic: the fundamental theory is
non-relativistic.

Perhaps it can be described as a Quantum Unification of All Non-relativistic Theories of the Universe and Matter
(QUANTUM). It makes only one prediction, namely that, in order for fundamental quantum physics
to be background-independent, as it obviously must be, the Standard Model must be background-dependent.
Subtle changes in the gravitational background will therefore cause subtle changes in experimental measurements,
particularly those that depend on measuring neutrino momenta. 
The Standard Model can be tweaked for each new anomaly, and surely will be,
to fit the new background. But 
the anomalies 
will keep on coming, until such
time as the background-dependence is recognised.

\end{document}